\documentclass[3p,times,authoryear,12pt,final]{elsarticle}
\usepackage{hyperref}
\usepackage[utf8]{inputenc}
\usepackage{graphicx}
\usepackage{amsmath}
\usepackage[table,xcdraw,dvipsnames]{xcolor}
\usepackage{booktabs}

\journal{Journal of Commodity Markets}
\begin{document}
\begin{frontmatter}
\title{Extrapolating the long-term seasonal component of electricity prices for forecasting in the day-ahead market}

\author[wroclaw]{Katarzyna Ch\c{e}\'c}
\author[wroclaw]{Bartosz Uniejewski}
\author[wroclaw]{Rafa{\l} Weron \corref{cor1}}
\address[wroclaw]{Department of Operations Research and Business Intelligence, Wrocław University of Science and Technology, 50-370 Wrocław, Poland}
\cortext[cor1]{Corresponding author; \textit{email:} rafal.weron@pwr.edu.pl}

\begin{abstract}
Recent studies provide evidence that decomposing the electricity price into the long-term seasonal component (LTSC) and the remaining part, predicting both separately, and then combining their forecasts can bring significant accuracy gains in day-ahead electricity price forecasting. However, not much attention has been paid to predicting the LTSC, and the last 24 hourly values of the estimated pattern are typically copied for the target day. To address this gap, we introduce a novel approach which extracts the trend-seasonal pattern from a price series extrapolated using price forecasts for the next 24 hours. We assess it using two 5-year long test periods from the German and Spanish power markets, covering the Covid-19 pandemic, the 2021/2022 energy crisis, and the war in Ukraine. Considering parsimonious autoregressive and LASSO-estimated models, we find that improvements in predictive accuracy range from 3\% to 15\% in terms of the root mean squared error and exceed 1\% in terms of profits from a realistic trading strategy involving day-ahead bidding and battery storage.
\end{abstract}

\begin{keyword}
Electricity price forecasting \sep Long-term seasonal component \sep Day-ahead market \sep Combining forecasts \sep Trading strategy
\end{keyword}
\end{frontmatter}

\section{Introduction}
\label{sec:Intro}

\let\thefootnote\relax\footnotetext{
	
Paper published in Journal of Commodity Markets 37, 100449 (2025)

DOI: 10.1016/j.jcomm.2024.100449}

Seasonal decomposition is a fundamental technique in time series analysis and forecasting whose origins can be traced back to the middle of the 19th century \citep{hyn:ath:21,pet:etal:22}. In its basic form, it decomposes the signal $P_{d,h}$ into three components: a deterministic slowly varying trend-cyclical component $T_{d,h}$ also called the \textit{long-term seasonal component} (LTSC), a deterministic regularly repeating pattern $s_{d,h}$ called the (short-term) seasonal component, and the remaining part, i.e., the ``residual'', also called the stochastic component $\eta_{d,h}$, in an additive fashion, i.e., $P_{d,h} = T_{d,h} + s_{d,h} + \eta_{d,h}$. Note that we use here the notation prevailing in the \textit{electricity price forecasting} (EPF) literature, especially in day-ahead forecasting, where $d$ refers to the day and $h$ to the hour of delivery \citep{wer:14}. 

Although commonly used in multi-step ahead forecasting, seasonal decomposition has not been extensively utilized in day-ahead EPF until \cite{now:wer:16} provided empirical evidence that it can be advantageous to decompose the electricity price $P_{d,h}$ into the deterministic LTSC and a stochastic component that includes the short-term seasonality, i.e., $Y_{d,h} = s_{d,h} + \eta_{d,h}$, predict both components separately, and then combine their forecasts: $\hat{P}_{d,h} = \hat{T}_{d,h} + \hat{Y}_{d,h}$. 
The rationale behind this approach is that the residuals obtained after seasonal decomposition better satisfy the assumptions underlying typical model architectures, including the regression-type models considered in this study. Interestingly, the approach has proven effective for both parsimonious autoregressive \cite[AR;][]{afa:fed:19,gro:nan:19,Shah2021,Zafar2022} and non-linear autoregressive neural network-type models \cite[NARX;][]{mar:uni:wer:19:narx}, as well as parameter-rich models estimated using the \textit{least absolute shrinkage and selection operator} \cite[LASSO;][]{jed:mar:wer:21}.
However, to our best knowledge, all existing studies in the context of day-ahead EPF have relied on a naive prediction of the LTSC, where the last 24 hourly values of the estimated trend-seasonal pattern were simply copied for the target day. 

To address this gap, we introduce a novel approach to extrapolating the LTSC for the next day. We show that extending the time series of electricity spot prices by the 24 hourly forecasts for the next day -- obtained with a base model that does not involve seasonal decomposition -- and then extracting the LTSC component for the 24 hours of the target day from the extended time series can lead to statistically significant accuracy gains compared to the naive approach used in the literature so far. 
We employ two techniques to extract the LTSC: a simple \textit{moving average} \cite[MA; see, e.g.,][]{wer:14} and \textit{wavelet smoothing} \cite[also called \textit{thresholding};][]{per:wal:00}. Moreover, we use averaging to capture the benefits of combining forecasts from different models \citep{pet:etal:24}. 

To allow for a thorough evaluation of the studied models, like \cite{wag:ram:sch:mic:22}, and in line with the best practices outlined in \cite{lag:mar:des:wer:21}, we use very long test sets from more than one market. Namely, we consider 9 years of data (2015-2023) from two major European power markets -- EPEX in Germany and OMIE in Spain. The out-of-sample test period spans 5 years (2019-2023), covering the Covid-19 pandemic and the 2021-2022 energy crisis with skyrocketing prices of electricity. The day-ahead electricity price forecasts are obtained either using a parsimonious autoregressive expert model with exogenous variables \cite[ARX;][]{bil:gia:del:rav:23,gai:gou:ned:16,mac:nit:wer:21,zie:wer:18} or a parameter-rich, LASSO-estimated regression model \cite[LEAR;][]{lag:mar:des:wer:21,wag:ram:sch:mic:22}. We assess the significance of differences in predictive performance using the multivariate variant of the Diebold-Mariano (DM) test, as introduced by \cite{zie:wer:18}. In line with a recent trend in the EPF literature \citep{mac:uni:wer:23}, we also consider a realistic trading strategy involving day-ahead bidding and battery storage in order to quantify the benefits in monetary terms. 

The remainder of the paper is structured as follows. In Section \ref{sec:data} we describe the datasets, then in Section \ref{sec:methodology} we present the methodology, i.e., describe the ARX and LEAR models, discuss the methods used to estimate and extrapolate the LTSC, and briefly explain the transformation used to stabilize the variance of the stochastic component. In Section \ref{sec:results} we discuss the main results. Finally, in Section \ref{sec:conclusions} we wrap up the findings and conclude.

\begin{figure*}[tbp]
\includegraphics[width = \textwidth]{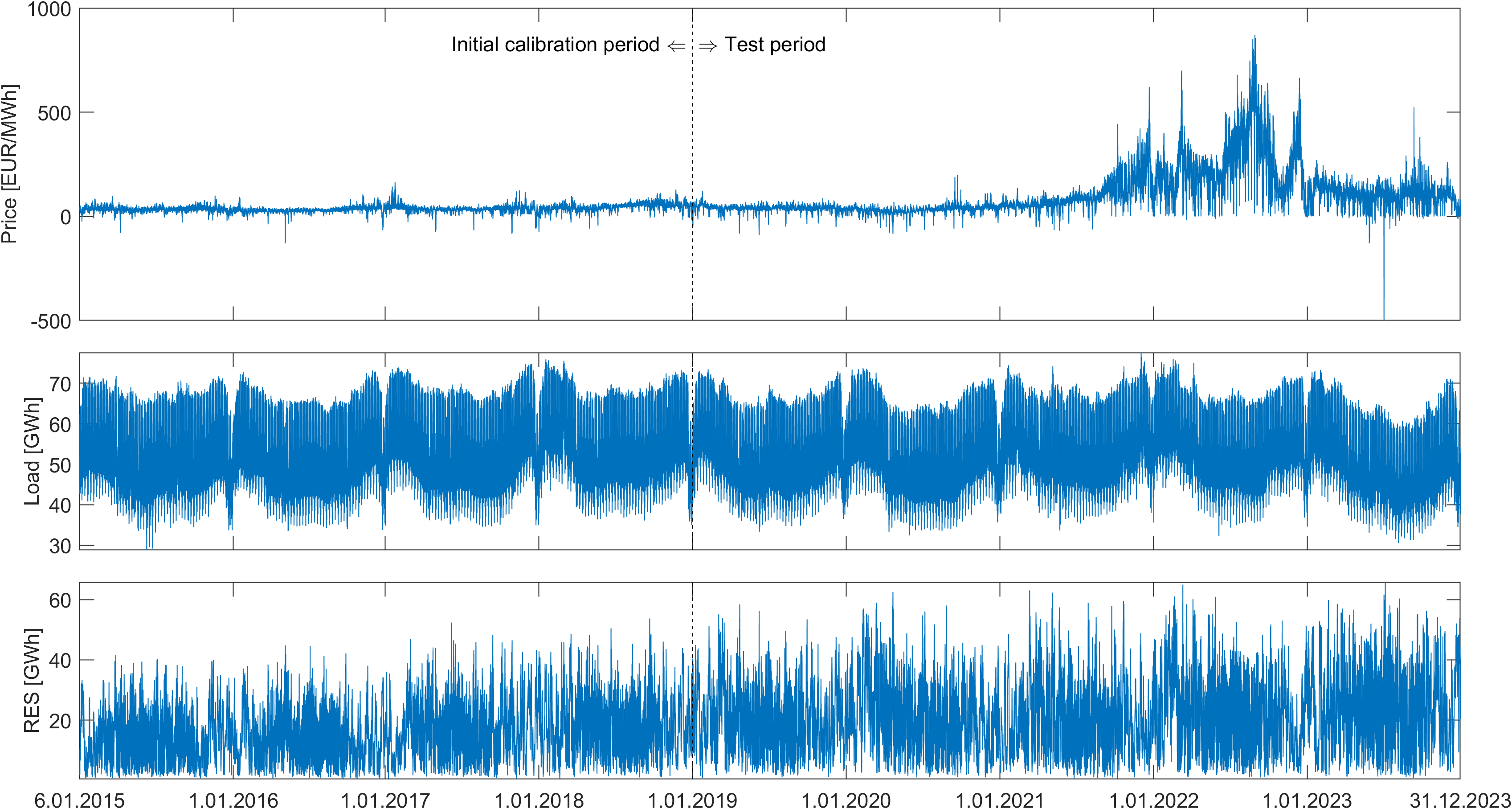}
\centering
\caption{Electricity prices (\textit{top}), day-ahead load forecasts (\textit{middle}) and day-ahead RES generation forecasts (\textit{bottom}) in the German electricity market (EPEX). The end of the initial calibration window is marked by the vertical dashed line.}
\label{fig:EPEX:208:data}
\end{figure*}

\begin{figure*}[tbp]
\includegraphics[width = \textwidth]{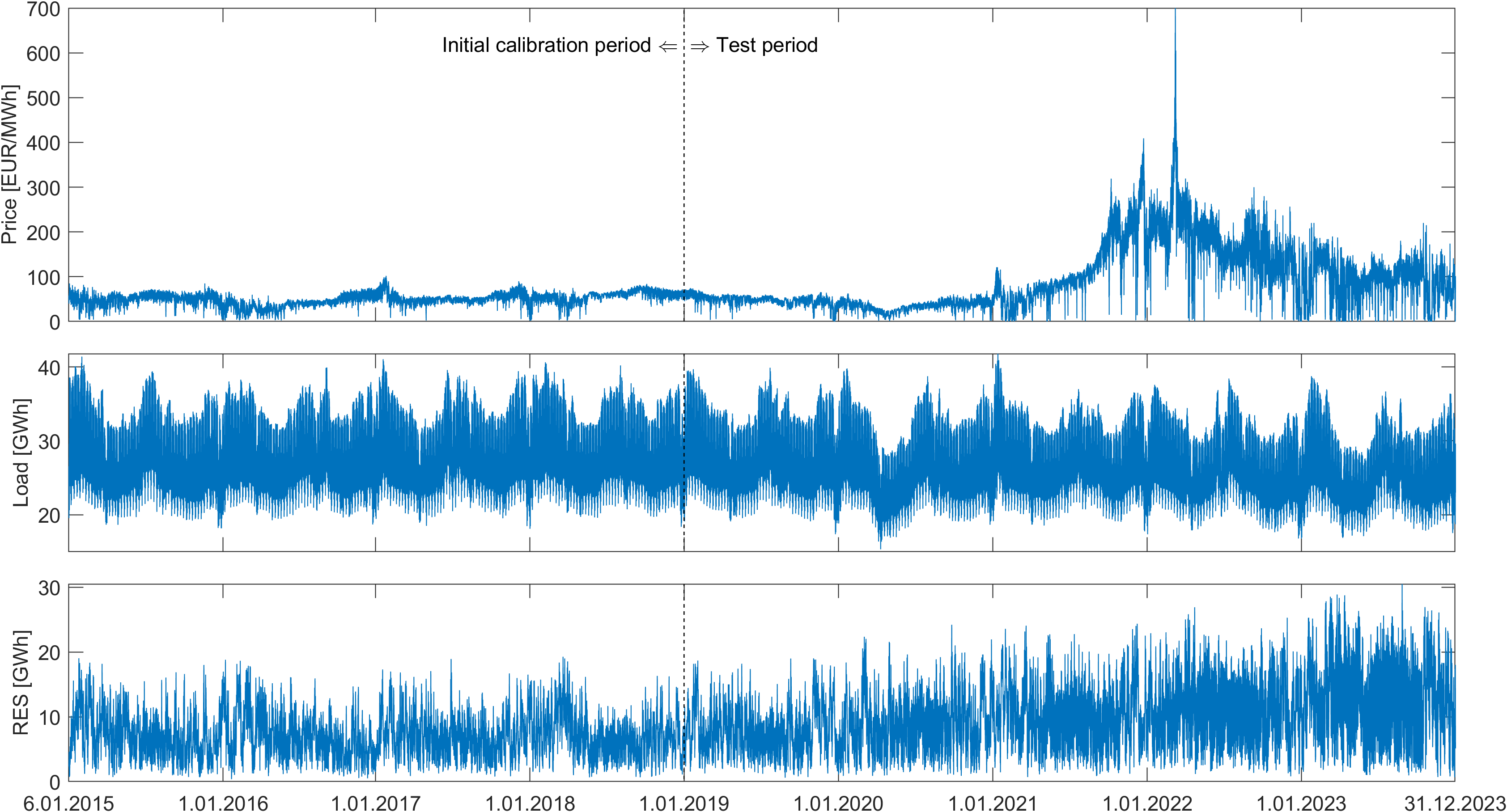}
\centering
\caption{Electricity prices (\textit{top}), day-ahead load forecasts (\textit{middle}) and day-ahead RES generation forecasts (\textit{bottom}) in the Spanish electricity market (OMIE). The end of the initial calibration window is marked by the vertical dashed line.}
\label{fig:OMIE:208:data}
\end{figure*}

\section{Data}
\label{sec:data}
 
We consider data from two major European power markets -- EPEX in Germany (see Figure \ref{fig:EPEX:208:data}) and OMIE in Spain (see Figure \ref{fig:OMIE:208:data}). The German market is one of the more studied ones, likely due to the central position in Europe, large wind and solar penetration amounting to over 45\% of net electricity generation in 2023 \citep{entsoe}, challenging price dynamics with abundant negative prices, and high trading liquidity in most segments \citep{cal:fus:ron:17,hag:etal:16,jan:24,kat:zie:18,mar:nar:wer:zie:23,nar:zie:20JCM,uni:24:ORD}. Although less studied in the literature \citep{dia:pla:16,g-m:car:san:15,gia:rav:ros:20,lip:uni:wer:24}, the Spanish market is very interesting because of the rapid increase in the share of solar generation -- from 5.1\% of net electricity generation in 2015 to 16.5\% in 2023 \citep{entsoe}.

Both datasets are of hourly resolution and include day-ahead electricity prices as well as forecasts of electricity consumption and generation from \textit{renewable energy sources} (RES; here solar and wind) downloaded from the ENTSO-E transparency platform  \citep{entsoe}. With the growing share of RES in the generation stack and the impact of its volume on electricity prices, day-ahead forecasts of solar and wind generation have become fundamental inputs to EPF models \citep{bil:gia:del:rav:23,jan:woj:22,wes:fle:etal:21}.
Both datasets have been preprocessed to handle missing/duplicate values that occur when switching to/from daylight saving time -- missing observations have been replaced by the arithmetic mean of two neighboring values, while duplicate ones have been replaced by their arithmetic mean \citep{wer:14}. To make the marginal distributions of prices and fundamental variables less leptokurtic and thus more suitable for modeling, we use the area hyperbolic sine (asinh) transformation \cite[][for details see Section \ref{ssec:VST}]{uni:wer:zie:18}. Note that unlike the logarithmic transformation that is popular in finance, the asinh can also handle negative values. 

Furthermore, in line with the EPF literature, we use a rolling window scheme \citep{gro:nan:19,lag:mar:des:wer:21,mar:ser:wer:18} to obtain day-ahead forecasts of electricity prices. Inspired by recent research which involved parameter-rich models \citep{mar:nar:wer:zie:23}, we use a calibration window of 208 weeks or 1456 days. To obtain electricity price forecasts for the first day of the test period, i.e., 1.01.2019, the models are calibrated to data spanning from 6.01.2015 to 31.12.2018. Next, the calibration window is moved forward one day and the forecasts are made for 2.01.2019. This process continues until electricity price forecasts are obtained for the last day in the test set, i.e., 31.12.2023.

\section{Methodology}
\label{sec:methodology}

In this study, we compute point forecasts of the 24 hourly prices for day $d$ using the information known at ca.\ 8 am on day $d-1$. Although the forecast horizon formally spans 16-40 hours, we actually make one step (day) ahead predictions in a multivariate modeling framework that uses a set of 24 interrelated models, one for each hour of day $d$ \citep{zie:wer:18}. 

More specifically, we use a multi-stage procedure. First, the electricity price is decomposed into the LTSC and a stochastic component that includes the short-term seasonality: $P_{d,h} = T_{d,h} + Y_{d,h}$. Then we apply a so-called \textit{variance stabilizing transformation} \cite[VST; see Section \ref{ssec:VST} and][]{uni:wer:zie:18} to $Y_{d,h}$ and obtain standardized series $y_{d,h}$. The latter is predicted using one of the models described in Section \ref{ssec:Models} to yield $\hat y_{d,h}$. In the last two stages, we apply the inverse VST and add the LTSC forecast to obtain the final price forecast $\hat P_{d,h}$:
\begin{equation}
\label{eq:procedure}
 P_{d,h} \xrightarrow[\text{Sec. \ref{ssec:LTSC:extracting}}]{-\text{LTSC}} Y_{d,h} \xrightarrow[\text{Sec. \ref{ssec:VST}}]{\text{VST}} y_{d,h} \xrightarrow[\text{Sec. \ref{ssec:Models}}]{\text{Predict}} \hat{y}_{d,h} \xrightarrow[\text{Sec. \ref{ssec:VST}}]{\text{VST}^{-1}} \hat{Y}_{d,h} \xrightarrow[\text{Sec. \ref{ssec:LTSC:forecasting}}]{+\widehat{\text{LTSC}}} \hat{P}_{d,h}.
\end{equation}
Note that the data are transformed after removing the LTSC and not vice versa, as this has been found to yield more accurate predictions in EPF \citep{jed:mar:wer:21}. Note also that for the benchmark models that do not use seasonal decomposition, we apply the VST directly to the raw price series, thus omitting the first and last stages of the above procedure.

\subsection{Forecasting with regression models}
\label{ssec:Models}

The day-ahead electricity price forecasts are obtained using two model classes. The first is a parsimonious autoregressive expert model with exogenous variables (ARX), originally proposed by \cite{mis:tru:wer:06}, later modified and compared in a number of EPF studies under different names and acronyms \citep{bil:gia:del:rav:23,gai:gou:ned:16,mac:nit:wer:21,mac:now:16,tay:21:expectile,zie:16:TPWRS,zie:wer:18}. In the ARX model the electricity price for day $d$ and hour $h$ is given by the following formula: 
\begin{equation}
\label{eq:ARX}
\begin{aligned}[b]
 y_{d,h}  = & \underbrace{\beta_1y_{d-1,h}+\beta_2y_{d-2,h}+\beta_3y_{d-7,h}}_\text{autoregressive effects} + \underbrace{\beta_4y_{d-1,24}}_\text{end-of-day} +\underbrace{\beta_5y_{d-1}^{min}+\beta_6y_{d-1}^{max}}_\text{non-linear effects} \\
 & + \underbrace{\beta_7X_{d,h}^1+\beta_8X_{d,h}^2}_\text{exogenous variables} + \underbrace{\sum\nolimits_{j=1}^7 \beta_{h,j+8}D_j}_\text{daily dummies} + ~\varepsilon_{d,h},
\end{aligned}
\end{equation}
where the first three regressors account for the autoregressive effects of the prices for the same hour on days $d-1$, $d-2$ and $d-7$, $y_{d-1,24}$ provides information on the last known price level, i.e., midnight of day $d-1$, $y^{max}_{d-1}$ and $y^{min}_{d-1}$ stand for the maximum and minimum price of the previous day, $X^1_{d,h}$ and $X^2_{d,h}$ are the exogenous variables -- respectively the day-ahead forecasts of the system-wide load and RES generation, $D_1,\dots,D_7$ are daily dummies, and $\varepsilon_{d,h}$ is the noise term. The coefficients $\beta_j$ are estimated using \textit{ordinary least squares} (OLS).

The second model class is a parameter-rich LASSO-estimated regression (LEAR) introduced to the EPF literature by \cite{uni:now:wer:16} and \cite{zie:16:TPWRS}, and later used by \cite{lag:rid:sch:18}, \cite{mac:uni:wer:23}, \cite{wag:ram:sch:mic:22} and \cite{zie:wer:18}, among others. Here, we consider a variant proposed by \cite{lag:mar:des:wer:21}, who coined the acronym LEAR:
\begin{equation}\label{eq:LEAR}
\begin{aligned}
y_{d,h} &= \underbrace{\sum\nolimits_{i=1}^{24} \left( \beta_{h,i} y_{d-1,i} 
+ \beta_{h,i+24} y_{d-2,i} + \beta_{h,i+48} y_{d-3,i}
+ \beta_{h,i+72}y_{d-7,i} \right)}_\text{autoregressive effects}  \\
&\quad + \underbrace{\sum\nolimits_{i=1}^{24} \left( \beta_{h,i+96} X^1_{d,i} + \beta_{h,i+120} X^1_{d-1,i} + \beta_{h,i+144} X^1_{d-7,i} \right)}_\text{exogenous variable \#1 and its lags}  \\
&\quad + \underbrace{\sum\nolimits_{i=1}^{24} \left( \beta_{h,i+168} X^2_{d,i} + \beta_{h,i+192} X^2_{d-1,i} + \beta_{h,i+216} X^2_{d-7,i} \right)}_\text{exogenous variable \#2 and its lags}  
+ \underbrace{\sum\nolimits_{j=1}^{7} \beta_{h,240+j} D_j}_\text{daily dummies}  + ~\varepsilon_{d,h},
\end{aligned}
\end{equation}
where the first 96 regressors are autoregressive terms that include prices from all hours of days $d-1$, $d-2$, $d-3$ and $d-7$, and the following $2\times 72$ regressors are all hourly values of the exogenous variables $X^1$ and $X^2$ for days $d$, $d-1$ and $d-7$.  
Such a model structure allows all cross-hour dependencies to be incorporated into the price forecasts. 
The LEAR model is estimated using the least absolute shrinkage and selection operator (LASSO) of \cite{tib:96}, which automatically selects the most relevant regressors for predicting $y_{d,h}$. 
Although many different regularization methods have been proposed in the literature, \cite{uni:24:ORD} identified LASSO as a parsimonious yet robust and well-performing variant in a comprehensive EPF evaluation study.

\subsection{Transforming the time series}
\label{ssec:VST}

Along with seasonal decomposition, data transformation is the primary preprocessing technique in time series analysis \citep{hyn:ath:21}. Its purpose is to remove known sources of variation, make the data more consistent across the sample, and -- particularly in EPF -- allow the handling of close to zero or negative values \citep{dia:pla:16,jan:24,kat:zie:18}. 
For comprehensive comparisons of different variance stabilizing transformations (VSTs) see \cite{cia:mun:zar:22}, \cite{shi:wan:che:ma:21} and \cite{uni:wer:zie:18}, and for a discussion of the order of applying seasonal decomposition and VSTs before model calibration see \cite{jed:mar:wer:21}.

Following \cite{lag:mar:des:wer:21} and \cite{zie:wer:18}, we use here the \textit{area hyperbolic sine} (asinh) transformation with the (median, MAD) normalization:
\begin{equation}
 y_{d,h} = \text{asinh}\left(\frac{Y_{d,h}-\textrm{Med}_\tau}{1.4826 \cdot \textrm{MAD}_\tau}\right), 
\end{equation}
where
\begin{equation}
 \text{asinh}(x) = \log\left(x+\sqrt{x^2+1}\right),
\end{equation}
$\text{Med}_\tau$ is the median and $\text{MAD}_\tau$ is the median absolute deviation of $Y_{d,h}$ in the calibration window $\tau$, and 1.4826 is the inverse of the 75th percentile of the standard normal distribution; this factor ensures asymptotic normal consistency with the standard deviation \citep{uni:wer:zie:18}.

Once $\hat{y}_{d,h}$ is computed, we apply the inverse transformation:  
\begin{equation}
\label{eq:back-trans}
 \hat{Y}_{d,h}=1.4826 \cdot \textrm{MAD}_\tau \cdot \textrm{sinh}\left(\hat{y}_{d,h}\right)+\textrm{Med}_\tau,
\end{equation}
where $\textrm{sinh}$ is the hyperbolic sine. As \cite{nar:zie:20JCM} note, the latter is not the correct inverse transformation since, given random variable $X$, $\mathbb{E}\textrm{sinh}(X)$ does not have to equal $\textrm{sinh}(\mathbb{E}X)$. Nevertheless, the difference between Eq.\ \eqref{eq:back-trans} and the correct inverse VST is not substantial and is generally ignored in the literature.

\subsection{Extracting the LTSC}
\label{ssec:LTSC:extracting}

When \cite{now:wer:16} introduced the seasonal component approach, they used two techniques for extracting the long-term seasonal component (LTSC): the Hodrick-Prescott filter \cite[HP; for sample applications in EPF see, e.g.,][]{cal:fus:ron:17,lis:nan:14,mar:uni:wer:19:narx,Zafar2022} and wavelet smoothing. Here, we replace the HP filter with a much simpler but equally effective simple moving average.

In the \textit{moving average} (MA) approach the LTSC is approximated by averaging observations within a window centered at $t=24d+h$:
\begin{equation}
 \hat{T}_{d,h} = \hat{T}_t = \frac{1}{m}\sum\nolimits_{j=-k}^{k} P_{t+j},
\end{equation}
where $m=2k+1$ is the width of the window. The latter affects the shape of the trend-seasonal component, i.e., low $m$ yields a more volatile series, while high $m$ smooths out the fluctuations. In our study, we consider five smoothing levels: 1, 7, 28, 56, and 91 days, corresponding to window sizes ranging from $m=24+1$ (about 1 day) to $m=91\cdot 24+1$ (about 3 months). This allows us to obtain models that react differently to sudden changes in market conditions. Such an approach is similar to averaging across calibration windows of different lengths in macroeconometrics \citep{pes:tim:07} or in EPF \citep{hub:mar:wer:19}.

On the other hand, \textit{wavelet smoothing} \cite[also called \textit{thresholding};][]{per:wal:00} applies the discrete wavelet transform to decompose the original series into a sum of the so-called approximation series $S_J$ capturing the general trend and a set of detail series $D_1,D_2,...,D_{J}$. At each step $j=1,...,J$ of this iterative procedure, details $D_j$ of a given frequency are removed to yield a smoother, but twice shorter signal $S_j$. In this study, we consider five smoothing levels $S_J$ with $J \in (5,7,9,10,11)$, corresponding to time scales ranging from $2^5=32$ hours (or ``1 day'') to $2^{11} = 2048$ hours (or ``3 months''). Similar to the MA approach, the use of different smoothing levels allows us to capture changes in market conditions. Following \cite{now:wer:16}, we use the Daubechies wavelet family of order 24. For sample applications of wavelet smoothing in EPF see \cite{afa:fed:19}, \cite{gro:nan:19}, \cite{jed:mar:wer:21} \cite{lis:nan:14} and \cite{mar:uni:wer:19:narx}, among others.

\subsection{Combining forecasts}
\label{ssec:combining}

The contemporary forecasting literature agrees that combining predictions from different models generally improves forecasting accuracy \citep{ati:20,pet:etal:22}. The same has been reported for energy forecasting \citep{hon:etal:20:OAJPE}, and EPF in particular \citep{ber:zie:24,hub:mar:wer:19,lag:mar:des:wer:21,nit:wer:23}. 
In this study, we use simple arithmetic averaging to combine forecasts obtained with (i) different levels of decomposition and (ii) different methods of decomposing the data. As a result, we consider three different classes of forecasting models and denote them by suffixes. The average the forecasts obtained for the five moving average windows is denoted by \textbf{--MA}. Similarly, the average of the forecasts obtained for the five different levels of wavelet smoothing is denoted by \textbf{--S}. Finally, the average of all 10 forecasts is denoted by \textbf{--MAS}. Although more sophisticated ensembling methods have been considered in the literature, the simple arithmetic average has been repeatedly shown to be robust and competitive  \citep{rav:bou:dij:15, mar:ser:wer:18, uni:mac:23}.

\begin{figure}[tb]
\includegraphics[width = 0.9\textwidth]{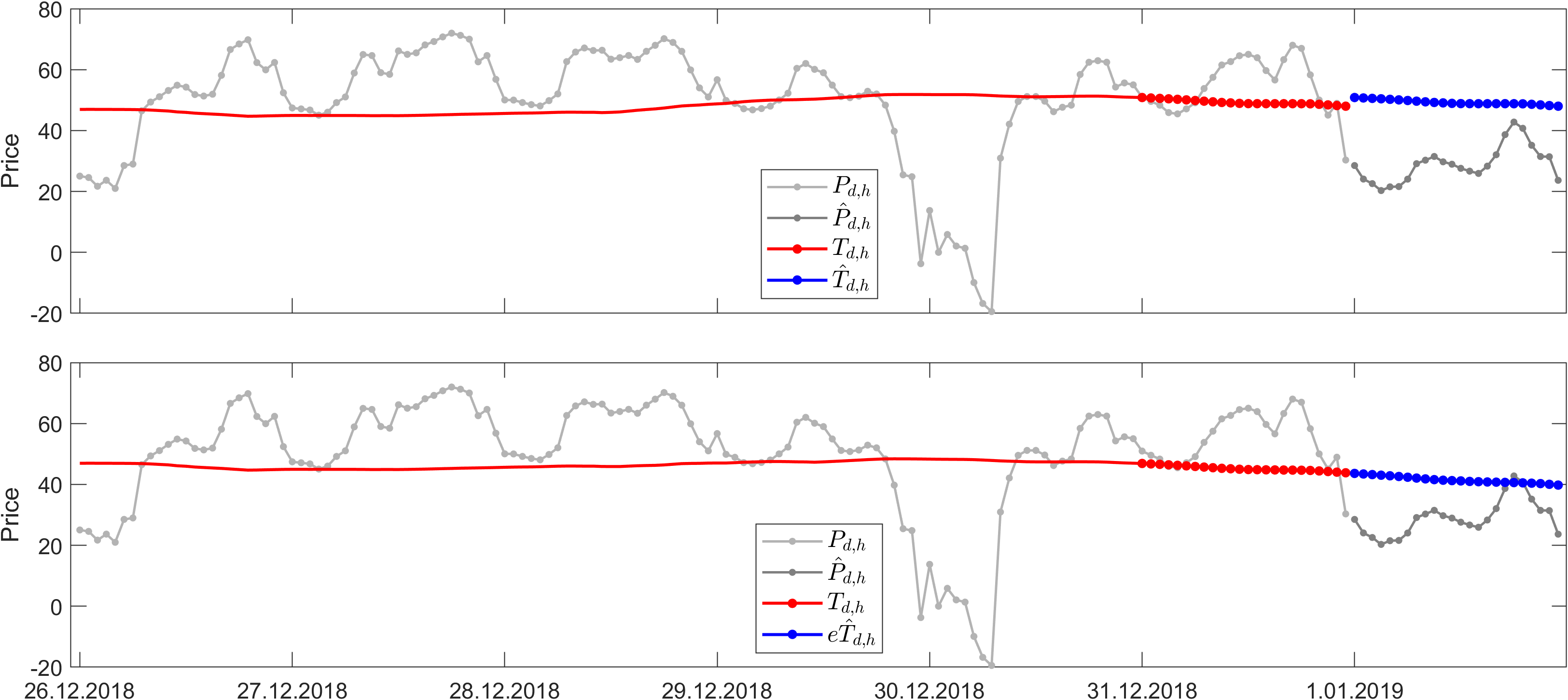}
\centering
\caption{Comparison of the naive LTSC day-ahead forecast ($\hat{T}_{d,h}$; \textit{top panel}) and the introduced in this study extrapolated LTSC forecast ($e\hat{T}_{d,h}$; \textit{bottom panel}). In both cases the LTSC is calculated using the moving average (MA) method with a 7-day smoothing window ($m=24\cdot7+1$).}
\label{fig:ltsc_day_ahead}
\end{figure}

\subsection{Extrapolating the LTSC}
\label{ssec:LTSC:forecasting}

To our best knowledge, all studies that use the seasonal component approach introduced by \cite{now:wer:16} rely on a naive prediction of the LTSC, where the last 24 hourly values of the estimated trend-seasonal pattern are simply copied for the target day. Depending on the pool of individual forecasts, we denote this approach by \textbf{SCARX}-$*$ or \textbf{SCLEAR}-$*$, where $*$ takes the value MA, S or MAS, see Section \ref{ssec:combining}, when referring to the model defined by Eq.\ \eqref{eq:ARX} or Eq.\ \eqref{eq:LEAR}, respectively.

In this paper, we introduce a new approach to predicting the LTSC for the next day by extrapolating the input price vector before calculating the LTSC. The proposed algorithm consists of the following steps:
\begin{enumerate}
 \item Compute the 24 hourly electricity price forecasts for the next day using the model without seasonal decomposition, i.e., ARX or LEAR.
 \item Append these forecasts to the prices in the calibration window, increasing the length of the calibration window by 24 observations.
 \item Use the extrapolated price vector as the input to one of the approaches to extracting the LTSC described in Section \ref{ssec:LTSC:extracting}.
\end{enumerate}
By following this procedure, we obtain the LTSC forecast for the target day along with the seasonal component of the prices in the calibration window. Therefore, this approach preserves the continuity of the LTSC vector, which transitions smoothly from the calibration period to the target day, see Figure \ref{fig:ltsc_day_ahead}. In our study, the seasonal component models that use the extrapolated LTSC approach are denoted by a prefix \textbf{e}: \textbf{eSCARX}-$*$ for ARX-based models and \textbf{eSCLEAR}-$*$ for LEAR-based models.

Note that we also considered a simpler method to extrapolate the LTSC, where the price series is extended for the next 24 hours using a naive forecast, i.e., $\hat{P}_{d,h} = P_{d-1,h}$, not an ARX or LEAR forecast as in steps 1-2 of the above algorithm. However, this approach resulted in inferior performance and its results are not reported here.

\section{Results}
\label{sec:results}

We evaluate the results using two metrics of prediction accuracy and a statistical test to assess the statistical significance of the results obtained \citep{lag:mar:des:wer:21,pet:etal:22}. For ensuring the reliability of our forecasts, we use 5-year long test periods, i.e., from 1.01.2019 to 31.12.2023, for both datasets. 

\subsection{Prediction accuracy}

We use the two most popular error metrics in the EPF literature \citep{cia:mun:zar:22,lag:mar:des:wer:21,mac:uni:wer:23,wer:14,zie:16:TPWRS} to measure the prediction accuracy of point forecasts, namely the Mean Absolute Error:
\begin{equation}\label{eq:MAE}
 \text{MAE} = \frac{1}{24N_d}\sum_{d=1}^{N_d}\sum_{h=1}^{24}|P_{d,h}-\hat{P}_{d,h}|,
\end{equation}
and the Root Mean Square Error (RMSE)
\begin{equation}\label{eq:RMSE}
 \text{RMSE} = \sqrt{\frac{1}{24N_d}\sum_{d=1}^{N_d}\sum_{h=1}^{24}(P_{d,h}-\hat{P}_{d,h})^2
 },
\end{equation}
where $N_d=1826$ stands for the number of days in the out-of-sample test period, i.e., 1.01.2019--31.12.2023, see Figures \ref{fig:EPEX:208:data} and \ref{fig:OMIE:208:data}, and $P_{d,h}$ and $\hat{P}_{d,h}$ respectively denote the actual and predicted price for day $d$ and hour $h$.

\subsubsection{Performance across the whole test period}

In Tables \ref{table:MAE_results} and \ref{table:RMSE_results} we report the MAE and RMSE errors for the considered models over the 5-year test period. Note, that cells are colored (red $\rightarrow$ high, green $\rightarrow$ low) independently for each market (EPEX, OMIE) and model class (ARX, LEAR). For instance, in Table \ref{table:MAE_results} the MAE of 11.309 is red because it is the highest (i.e., worst) value among LEAR-class models for the OMIE dataset. In both tables columns labeled ``\%chng.'' show the percentage difference between the error (MAE or RMSE) for an extrapolated LTSC-type model (eSCARX, eSCLEAR) and a naive LTSC-type model (SCARX, SCLEAR).
Recall from Section \ref{ssec:combining} that suffixes -MA, -S and -MAS denote combined predictions for three pools of individual forecasts.

\begin{table*}[tb]
\caption{Mean Absolute Errors (MAE) for the considered models over the 5-year test period (01.01.2019--31.12.2023). Cells are colored (red $\rightarrow$ high, green $\rightarrow$ low) independently for each market (EPEX, OMIE) and model class (ARX, LEAR). Columns labeled ``\%chng.'' show the percentage difference between the MAE for an eSC-type model and an SC-type model in the two columns on the left.}
\label{table:MAE_results}
\begin{center}
\resizebox{\textwidth}{!}{
\begin{tabular}{lcccccccccccl}
\hline
& \multicolumn{6}{c}{EPEX} & \multicolumn{6}{c}{OMIE} \\ 
LTSC  & \multicolumn{2}{c}{ARX} & & \multicolumn{2}{c}{LEAR} 
& & \multicolumn{2}{c}{ARX} & & \multicolumn{2}{c}{LEAR} \\
None& \multicolumn{2}{c}{\cellcolor[HTML]{FFEB84}17.849}  & & \multicolumn{2}{c}{\cellcolor[HTML]{FFEB84}14.761}  &  & \multicolumn{2}{c}{\cellcolor[HTML]{F8696B}12.056}  & & \multicolumn{2}{c}{\cellcolor[HTML]{F8696B}11.309}  &\\ \hline
& SCARX & eSCARX & \%chng. & SCLEAR & eSCLEAR & \%chng. 
& SCARX & eSCARX & \%chng. & SCLEAR & eSCLEAR & \%chng. \\
--MA  & \cellcolor[HTML]{F8696B}18.373 & \cellcolor[HTML]{79C47C}17.011 & $-$7.70\%  & \cellcolor[HTML]{FA8A72}15.809 & \cellcolor[HTML]{94CC7D}14.116 & $-$11.33\%  & \cellcolor[HTML]{FDC47D}11.706 & \cellcolor[HTML]{BBD780}11.373 & $-$2.89\%  & \cellcolor[HTML]{FDBD7C}11.029 & \cellcolor[HTML]{9ACD7E}10.416 & \multicolumn{1}{c}{$-$5.71\%} \\
--S& \cellcolor[HTML]{FB9975}18.182 & \cellcolor[HTML]{A7D17E}17.297 & $-$4.99\%  & \cellcolor[HTML]{F8696B}16.155 & \cellcolor[HTML]{8BC97D}14.062 & $-$13.87\%  & \cellcolor[HTML]{FFE884}11.569 & \cellcolor[HTML]{63BE7B}11.130 & $-$3.87\%  & \cellcolor[HTML]{FFE082}10.914 & \cellcolor[HTML]{63BE7B}10.167 & \multicolumn{1}{c}{$-$7.09\%} \\
--MAS & \cellcolor[HTML]{FDBC7B}18.041 & \cellcolor[HTML]{63BE7B}16.870 & $-$6.71\%  & \cellcolor[HTML]{FB9273}15.717 & \cellcolor[HTML]{63BE7B}13.820 & $-$12.86\%  & \cellcolor[HTML]{FFEB84}11.556 & \cellcolor[HTML]{6EC17B}11.163 & $-$3.46\%  & \cellcolor[HTML]{FFEB84}10.874 & \cellcolor[HTML]{63BE7B}10.167 & \multicolumn{1}{c}{$-$6.72\%} \\
\hline
\end{tabular}
}
\end{center}
\end{table*}

\begin{table*}[tb]
\caption{Root Mean Square Errors (RMSE) for the considered models over the 5-year test period (01.01.2019--31.12.2023). Cells are colored (red $\rightarrow$ high, green $\rightarrow$ low) independently for each market (EPEX, OMIE) and model class (ARX, LEAR). Columns labeled ``\%chng.'' show the percentage difference between the RMSE for an eSC-type model and an SC-type model in the two columns on the left.}
\label{table:RMSE_results}
\begin{center}
\resizebox{\textwidth}{!}{
\begin{tabular}{lcccccccccccl}
\hline
\multicolumn{1}{r}{} & \multicolumn{6}{c}{EPEX} & \multicolumn{6}{c}{OMIE}  \\
LTSC  & \multicolumn{2}{c}{ARX}  && \multicolumn{2}{c}{LEAR} & & \multicolumn{2}{c}{ARX}  && \multicolumn{2}{c}{LEAR} &  \\
None  & \multicolumn{2}{c}{\cellcolor[HTML]{F8E983}31.531}  && \multicolumn{2}{c}{\cellcolor[HTML]{F2E783}26.861}  & & \multicolumn{2}{c}{\cellcolor[HTML]{FA8871}20.409}  && \multicolumn{2}{c}{\cellcolor[HTML]{F8696B}19.690}  &  \\ \hline
& SCARX  & eSCARX & \%chng. & SCLEAR & eSCLEAR& \%chng.  & SCARX  & eSCARX & \%chng. & SCLEAR & eSCLEAR& \%chng.\\
--MA  & \cellcolor[HTML]{F8696B}34.352 & \cellcolor[HTML]{FFEB84}31.555 & $-$8.49\% & \cellcolor[HTML]{FA7E6F}30.112 & \cellcolor[HTML]{FFEB84}26.927 & $-$11.18\% & \cellcolor[HTML]{F8696B}20.472 & \cellcolor[HTML]{AAD27F}19.810 & $-$3.29\% & \cellcolor[HTML]{F9726D}19.660 & \cellcolor[HTML]{A3D07E}18.616 & \multicolumn{1}{c}{$-$5.45\%} \\
--S& \cellcolor[HTML]{FA8671}33.735 & \cellcolor[HTML]{DEE182}31.443 & $-$7.04\% & \cellcolor[HTML]{F8696B}30.717 & \cellcolor[HTML]{9ECF7E}26.436 & $-$15.01\% & \cellcolor[HTML]{FECE7F}20.267 & \cellcolor[HTML]{63BE7B}19.471 & $-$4.00\% & \cellcolor[HTML]{FFDA81}19.299 & \cellcolor[HTML]{64BE7B}18.192 & \multicolumn{1}{c}{$-$5.91\%} \\
--MAS & \cellcolor[HTML]{FA8E73}33.567 & \cellcolor[HTML]{63BE7B}31.008 & $-$7.93\% & \cellcolor[HTML]{FA8671}29.877 & \cellcolor[HTML]{63BE7B}26.130 & $-$13.40\% & \cellcolor[HTML]{FFEB84}20.207 & \cellcolor[HTML]{69BF7B}19.504 & $-$3.54\% & \cellcolor[HTML]{FFEB84}19.239 & \cellcolor[HTML]{63BE7B}18.179 & \multicolumn{1}{c}{$-$5.66\%} \\ \hline
\end{tabular}
}
\end{center}
\end{table*}

Several important conclusions can be drawn:
\begin{itemize}
\setlength{\itemsep}{0pt}
\setlength{\parskip}{0pt}
 \item All extrapolated LTSC-based models (eSCARX, eSCLEAR) beat their naive LTSC counterparts (SCARX, SCLEAR) by a wide margin.
 \item The extrapolated LTSC-based models (eSCARX, eSCLEAR) are able to outperform the corresponding benchmark models without seasonal decomposition. 
 \item The moving average (MA) approach is confirmed to be a very useful tool for identifying the LTSC. 
 \item The LEAR-based models always outperform the corresponding ARX-based models.
\end{itemize}

In more detail, for models based on the parsimonious ARX structure, the reductions range from 2.89\% to 7.70\% in MAE and from 3.29\% to 8.49\% in RMSE. For the parameter-rich LEAR-type models, the improvements are even more striking -- from 5.71\% to 13.87\% in MAE and from 5.45\% to 15.01\% in RMSE. This shows the high effectiveness of the method introduced in this paper compared to the naive approach used so far. This is especially true for the German EPEX market, where much higher improvements in accuracy are observed, see the 7th column in Tables \ref{table:MAE_results} and \ref{table:RMSE_results}. Interestingly, the improvements with respect to the classical ARX and LEAR models that do not use seasonal decomposition are higher for the Spanish OMIE market and reach up to 10.65\% for SCLEAR-S and -MAS models in terms of the MAE and 7.98\% for the SCLEAR-MAS model in terms of the RMSE; not reported in the Tables. Apparently, the SCARX and SCLEAR models, which naively extrapolate the LTSC for the next day, cannot cope with the extreme electricity price dynamics in Germany over the studied period, see Figure \ref{fig:EPEX:208:data}.


In the case of the EPEX dataset, the best performing averaging approach is -MAS, which combines all MA- and wavelet-based forecasts. It is worth noting that only the extrapolated LTSC approach is able to outperform the corresponding benchmarks in Germany, as the SC-type models provide less accurate price forecasts. For the OMIE market, all three averaging schemes -MA, -S, and -MAS show good performance. This time, both the naive and the extrapolated LTSC approaches outperform the corresponding benchmarks, although the latter are more accurate.

\begin{figure*}[tb]
\centering
\includegraphics[width=0.4\textwidth]{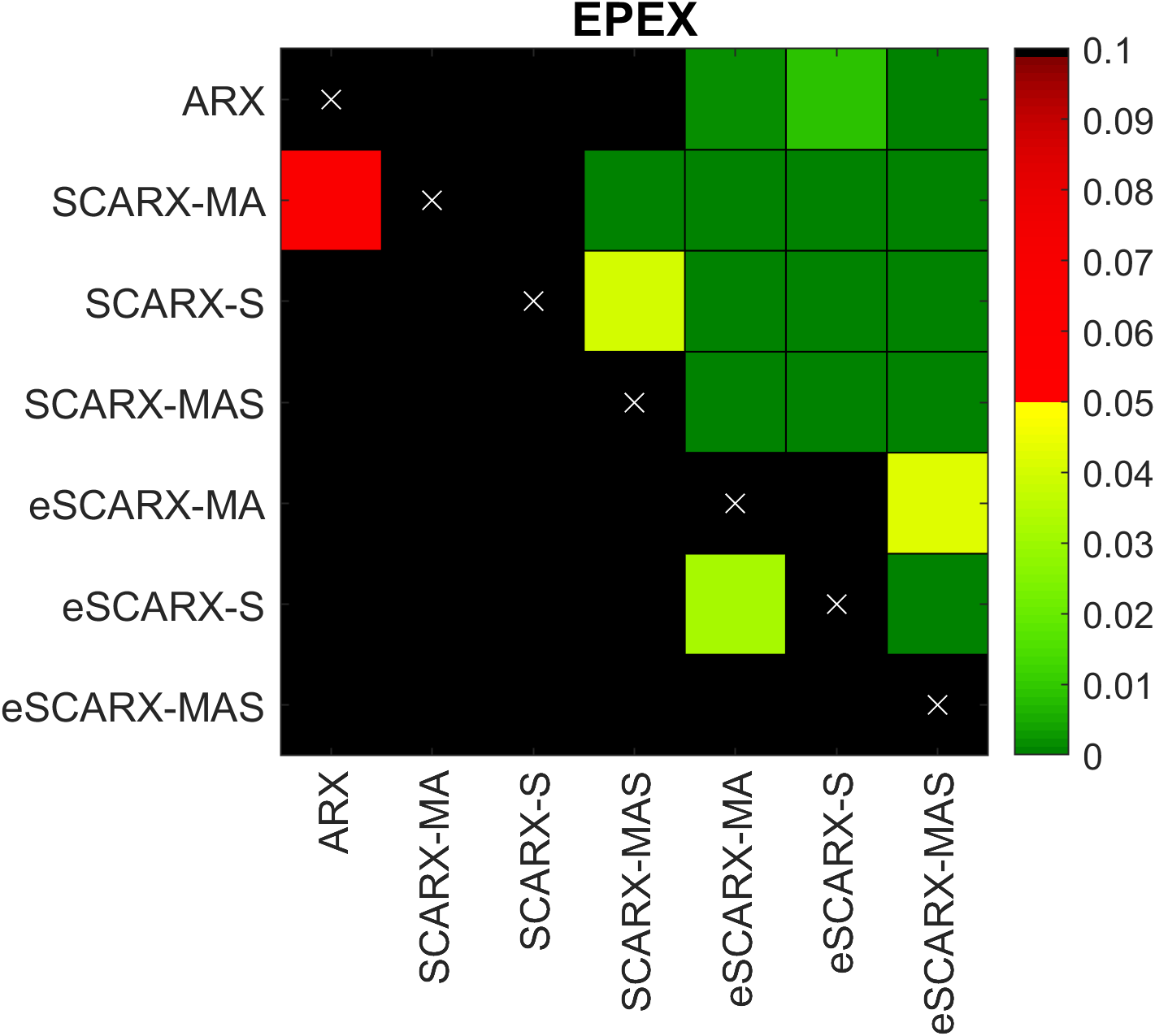}
\includegraphics[width=0.4\textwidth]{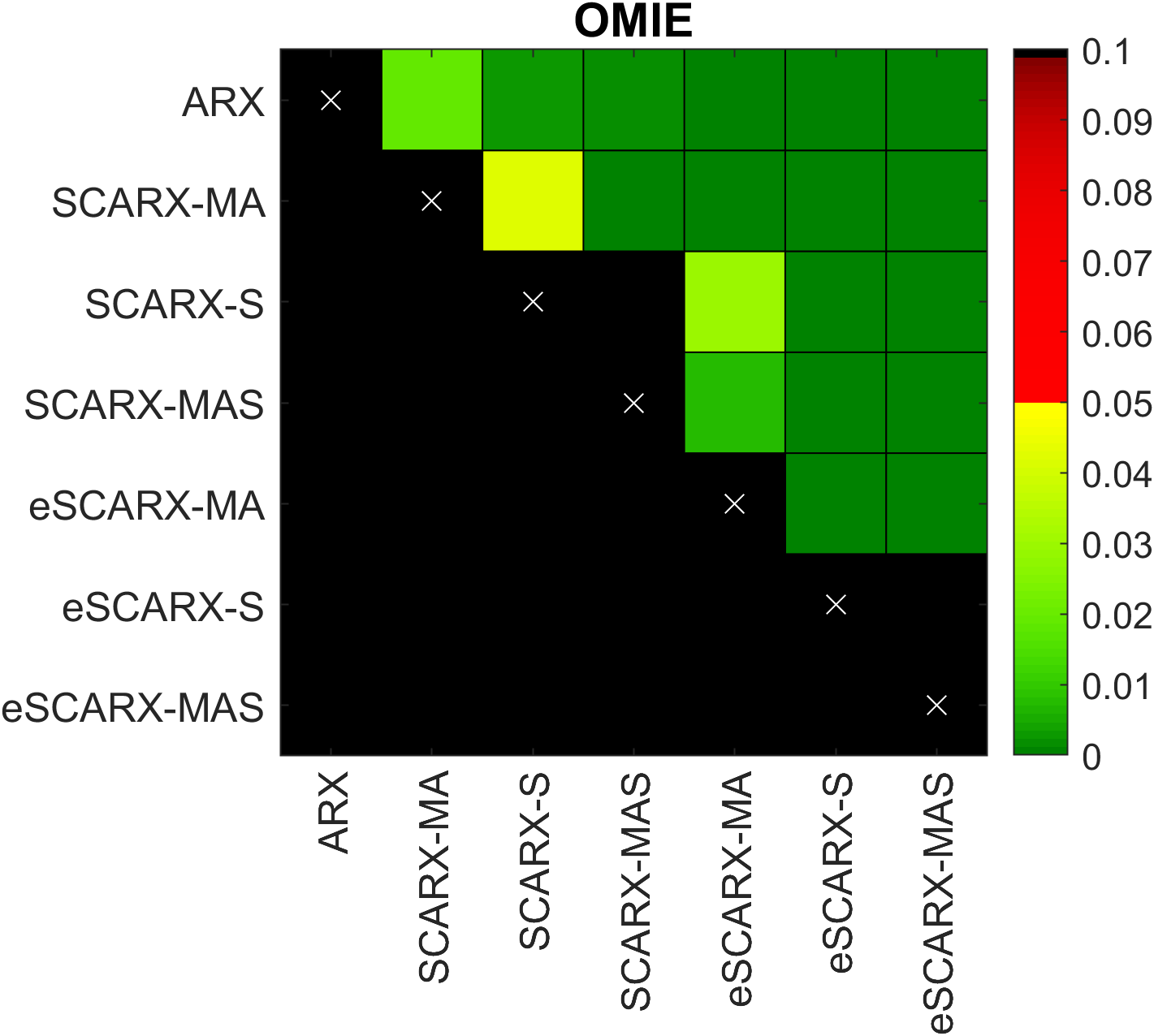}
\includegraphics[width=0.4\textwidth]{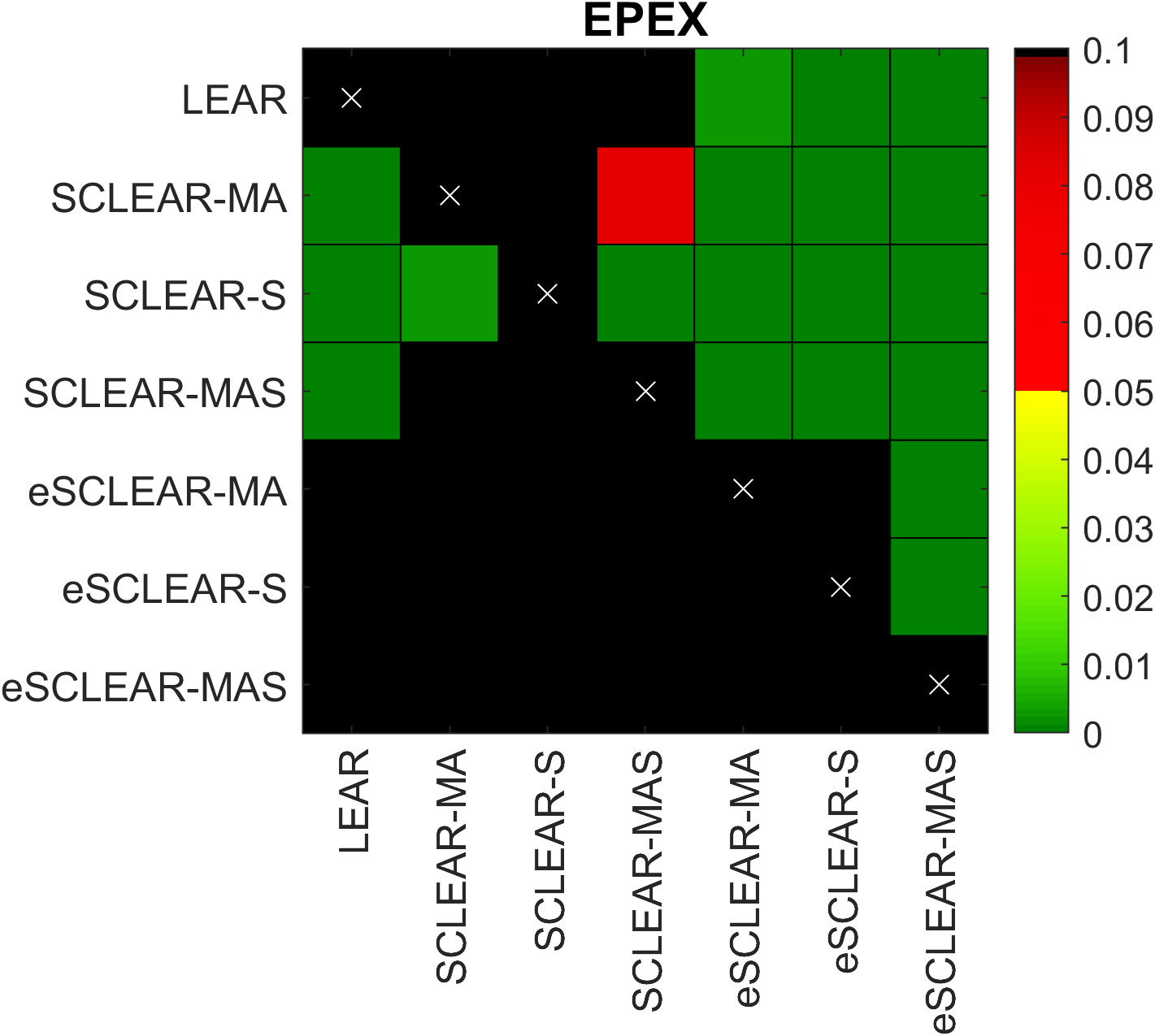}
\includegraphics[width=0.4\textwidth]{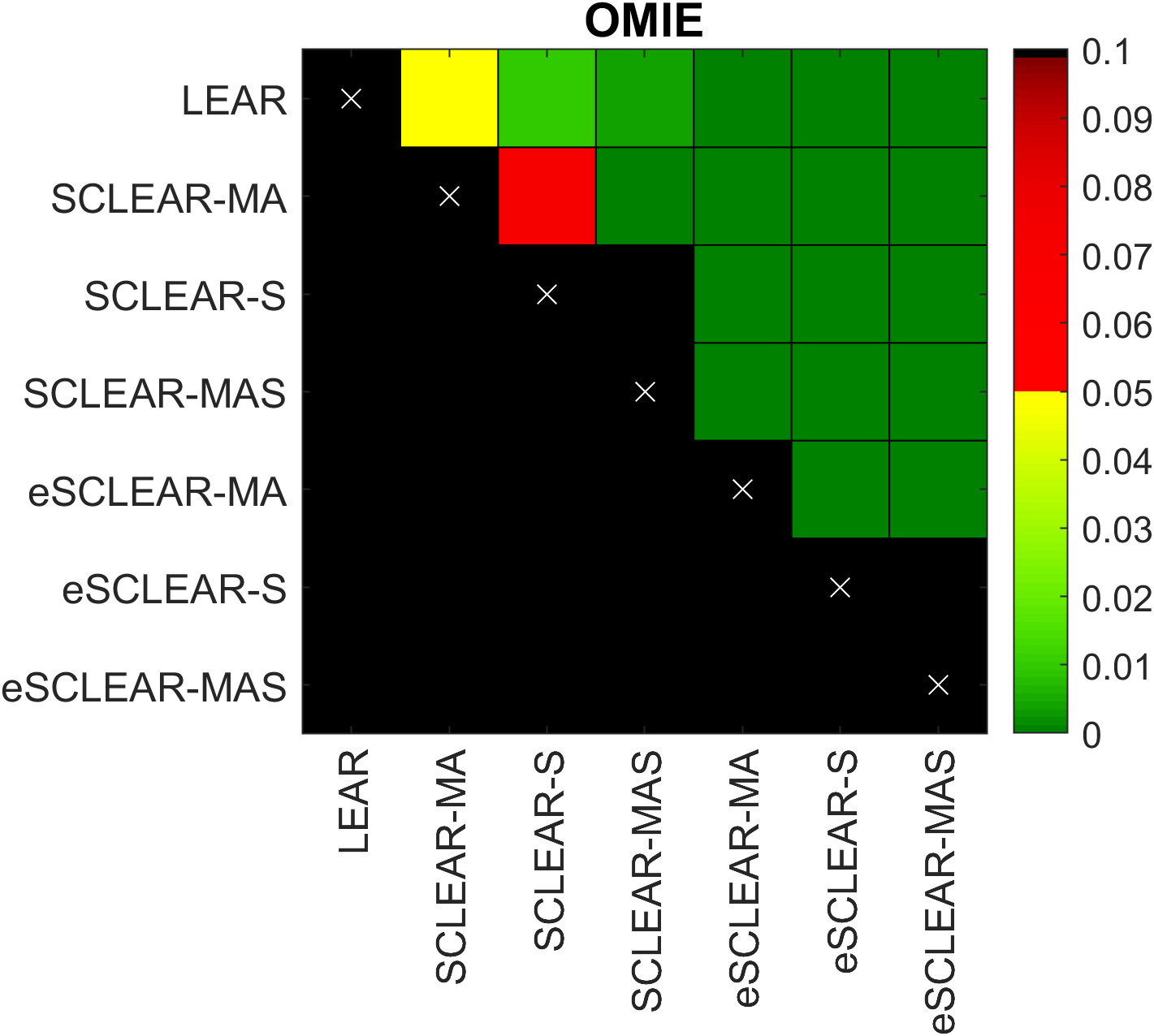}
\caption{Results of the Diebold-Mariano (DM) test with the multivariate loss differential series \eqref{eq:DM} and the $||.||_1$ norm, separately for the two datasets (EPEX -- \textit{left}, OMIE -- \textit{right}) and model classes (ARX -- \textit{top}, LEAR -- \textit{bottom}).
Like in \cite{lag:mar:des:wer:21} and \cite{zie:wer:18}, we use a heat map to indicate the range of the $p$-values. The closer they are to zero (dark green) the more significant is the difference between the forecasts of a model on the $X$-axis (better) and the forecasts of a model on the $Y$-axis (worse); $p$-values $\ge 0.10$ are marked in black.}
\label{fig:EPEX:OMIE:LEAR:DM}
\end{figure*}

\subsubsection{Statistical significance}
\label{sssec:results:DM}

Following \cite{zie:wer:18}, we use the multivariate variant of the \cite{die:mar:95} test (DM) to assess the statistical significance of differences in predictive performance. This version of the DM test is an asymptotic \textit{z}-test of the hypothesis that the mean of the `daily' or `multivariate' loss differential series \citep{mac:uni:wer:23,nar:zie:20JCM}:
\begin{equation}\label{eq:DM}
 \Delta_d^{A,B} = ||\varepsilon_d^A||_r - ||\varepsilon_d^B||_r,
\end{equation}
is zero, where $||\varepsilon_{d}^Z||_r = (\sum_{h=1}^{24} |\varepsilon_{d,h}^Z|^r)^{1/r}$ is the $r$-th norm of the 24-dimensional vector $\varepsilon_d^Z$ of out-of-sample errors for model $Z$. In our study, we use the $||.||_1$ norm, i.e., set $r=1$. Naturally, we assume that the loss differential series is covariance stationary.

In Figure \ref{fig:EPEX:OMIE:LEAR:DM} we present the results using four chessboards, separately for the two datasets (EPEX, OMIE) and the two model classes (ARX, LEAR). Like in \cite{lag:mar:des:wer:21} and \cite{zie:wer:18}, each chessboard uses a heat map to indicate the range of the $p$-values: green and yellow for $p<0.05$, red for $0.05\le p <0.10$, and black for $0.10 \le p$.

For both markets, the eSCARX-based models significantly outperform the ARX, and similarly, the eSCLEAR-based models outperform the LEAR benchmark. On the other hand, only in the case of the OMIE market do the SCARX- and SCLEAR-based models provide more accurate forecasts than the respective benchmark models, but at a lower level of significance.

Furthermore, all models with extrapolated LTSC significantly outperform the corresponding naive LTSC-based models and for all averaging schemes, i.e., -MA, -S, and -MAS.  In the majority of cases, the -MAS approach outperforms both the -MA and -S averaging schemes, which is not surprising since -MAS models use a larger and more diverse pool of individual forecasts. Overall, the eSCLEAR-MAS model is significantly better than any of the other models, with the exception of eSCLEAR-S for the OMIE market, but even in that case it is not significantly worse.

\begin{figure*}[tb]
\includegraphics[width = \textwidth]{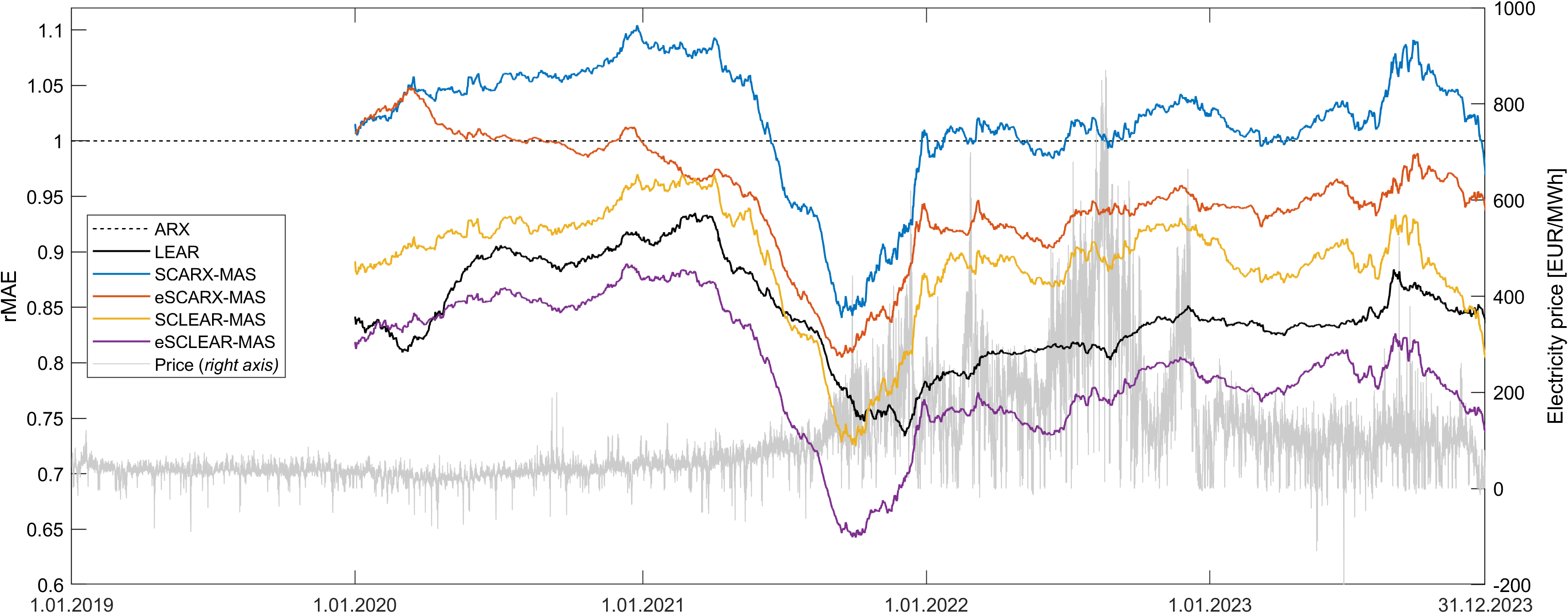}
\centering
\caption{Rolling 365-day \textit{relative mean absolute errors} (rMAE; \textit{left axis}) for the German (EPEX) market with respect to MAE of the ARX model, see Eq.\ \eqref{eq:rMAE}. Values for 1.01.2020 are the relative MAEs for 1.01.2019--31.12.2019, values for 2.01.2020 are the relative MAEs for 2.01.2019--1.01.2020, etc. The gray curve is the mean daily electricity price (in EUR/MWh; \textit{right axis}) in the depicted period.}
\label{fig:EPEX:208}
\end{figure*}

\begin{figure*}[tb]
\includegraphics[width = \textwidth]{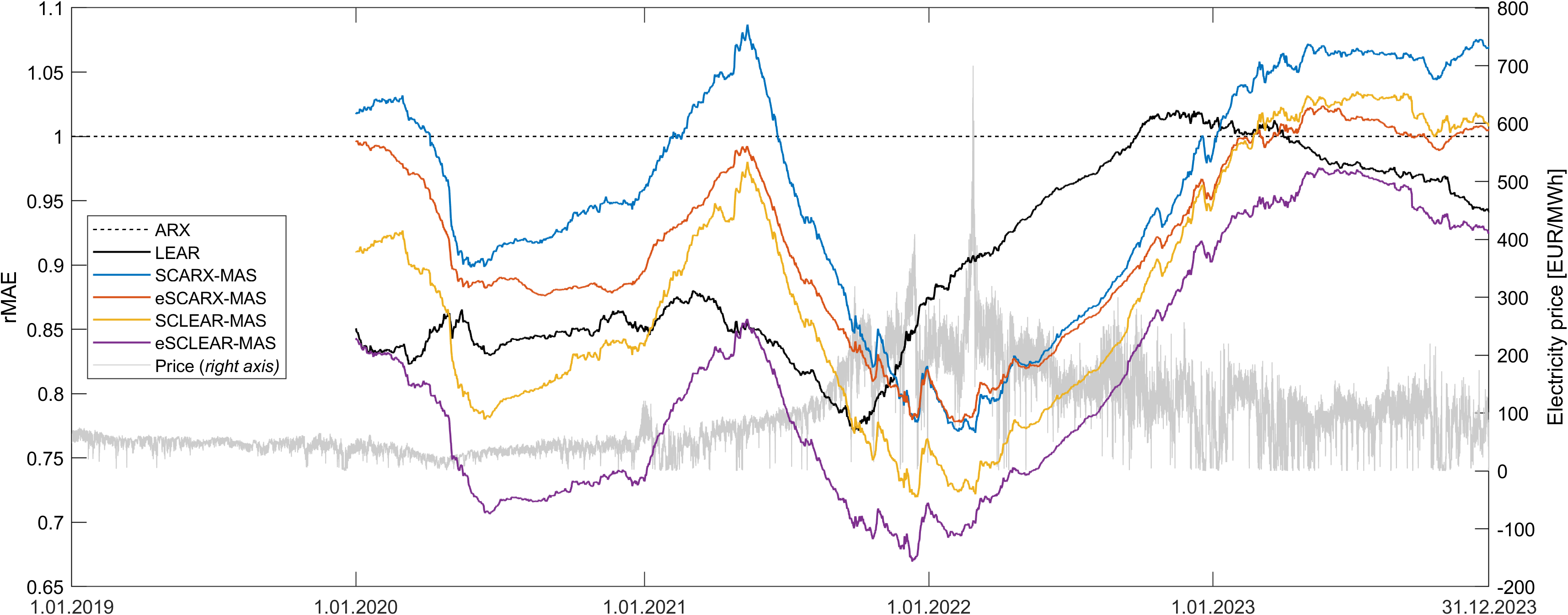}
\centering
\caption{Rolling 365-day \textit{relative mean absolute errors} (rMAE; \textit{left axis}) for the Spanish (OMIE) market with respect to MAE of the ARX model, see Eq.\ \eqref{eq:rMAE}. Values for 1.01.2020 are the relative MAEs for 1.01.2019--31.12.2019, values for 2.01.2020 are the relative MAEs for 2.01.2019--1.01.2020, etc. The gray curve is the mean daily electricity price (in EUR/MWh; \textit{right axis}) in the depicted period.}
\label{fig:OMIE:208}
\end{figure*}

\subsubsection{Temporal performance}
\label{sssec:Temporal}

In Figures \ref{fig:EPEX:208} and \ref{fig:OMIE:208} we plot the rolling 365-day \textit{relative mean absolute errors} (rMAE) for the best performing models, i.e., based on the richest pool of individual forecasts labeled --MAS, with respect to the error of the ARX model:
\begin{equation}\label{eq:rMAE}
\text{rMAE}(\delta) 
= \frac{\text{MAE}_{model}(\delta)}{\text{MAE}_{\text{ARX}}(\delta)} 
= \frac{\frac{1}{24\cdot 365}\sum_{d=\delta-365}^{\delta-1}\sum_{h=1}^{24}|P_{d,h}-\hat{P}_{model,d,h}|}{\frac{1}{24\cdot 365}\sum_{d=\delta-365}^{\delta-1}\sum_{h=1}^{24}|P_{d,h}-\hat{P}_{\text{ARX},d,h}|},
\end{equation}
where $P_{d,h}$, $\hat{P}_{model,d,h}$ and $\hat{P}_{\text{ARX},d,h}$ respectively denote the actual price, its $model$-derived forecast and its ARX-derived forecast for day $d$ and hour $h$, and $\delta=1.01.2020, 2.01.2020, ..., 31.12.2023$. Note that if $model = \text{ARX}$ then $\text{rMAE} \equiv 1$, see the dashed horizontal lines. Values of rMAE below 1 indicate better performance than that of the benchmark ARX model, while values above 1 indicate worse performance.

Several important conclusions can be drawn from Figures \ref{fig:EPEX:208} and \ref{fig:OMIE:208}:
\begin{itemize}
\setlength{\itemsep}{0pt}
\setlength{\parskip}{0pt}
    \item Throughout the test period, the eSC-type models (red and purple lines) achieve higher accuracy than the corresponding SC-type models (blue and yellow lines). This is true for both ARX-based and LEAR-based models.
    \item SCARX-MAS (blue line) is the worst performing model for most of the time, being outperformed not only by other SC-type models, but also sometimes by the ARX benchmark (dashed horizontal line).
    \item The best performing model is the eSCLEAR-MAS (purple line) for both the EPEX and OMIE markets. It is the only one that consistently outperforms the LEAR benchmark, except for a very brief period (365-day windows ending in February and March 2020) in Germany.
    \item In the case of the EPEX market, the overall ranking of the forecasting models barely changes over time. 
    \item For OMIE market the performance of the LEAR benchmark changes rapidly over time. It outperforms all models except eSCLEAR-MAS until the end of 2021, then it provides one of the less accurate predictions in the year 2022 and regains the second place in 2023.
\end{itemize}

\subsection{Economic evaluation}
\label{ssec:results:profits}

To assess the practical value of reducing price forecasting errors for decision-makers, we now consider a realistic trading strategy in the day-ahead market. More specifically, we assume that we own a 1.25 MWh battery, which for technical reasons cannot be discharged below 0.25 MWh (or 20\% of the nominal capacity) and its efficiency of charging as well as discharging is ca.\ 90\% \cite[which corresponds to a ca.\ 80\% efficiency for a single charge-discharge cycle;][]{sik:etal:19}. 
Originally proposed by \cite{uni:wer:21}, the strategy involves placing a bid to buy electricity when prices are low and charge the battery, and simultaneously placing a bid to sell electricity when prices are high after discharging the battery; see also \cite{mar:nar:wer:zie:23} and \cite{nit:wer:23} who used variants of this strategy. 

\begin{table*}[tb]
\caption{Trading profits for the entire out-of-sample test period expressed as fractions of the profit for the crystal ball strategy, i.e., $\Pi_{\text{CB}} = 110~897.27$ EUR for EPEX and $\Pi_{\text{CB}} = 55~236.55$ EUR for OMIE. Columns labeled ``\%chng.'' show the percentage difference between the profit for an eSC-type model and an SC-type model in the two columns on the left.}
\label{table:profits}
\begin{center}
\resizebox{\textwidth}{!}{
\begin{tabular}{lcclcclcclccl}
\hline
  & \multicolumn{6}{c}{EPEX}  & \multicolumn{6}{c}{OMIE}  \\
LTSC & \multicolumn{2}{c}{ARX} &  & \multicolumn{2}{c}{LEAR}&  & \multicolumn{2}{c}{ARX} &  & \multicolumn{2}{c}{LEAR}&  \\
None & \multicolumn{2}{c}{\cellcolor[HTML]{FDC87D}0.881}&  & \multicolumn{2}{c}{\cellcolor[HTML]{89C97E}0.888}&  & \multicolumn{2}{c}{\cellcolor[HTML]{B7D780}0.830}&  & \multicolumn{2}{c}{\cellcolor[HTML]{F8696B}0.794}&  \\ \hline
  & SCARX& eSCARX  & \multicolumn{1}{c}{\%chng.} & SCLEAR  & eSCLEAR & \multicolumn{1}{c}{\%chng.} & SCARX& eSCARX  & \multicolumn{1}{c}{\%chng.} & SCLEAR  & eSCLEAR & \multicolumn{1}{c}{\%chng.} \\
--MAS  & \cellcolor[HTML]{F8696B}0.876 & \cellcolor[HTML]{C9DC81}0.886 & \multicolumn{1}{c}{1.08\%}  & \cellcolor[HTML]{FBA977}0.880 & \cellcolor[HTML]{63BE7B}0.890 & \multicolumn{1}{c}{1.07\%}  & \cellcolor[HTML]{FEE482}0.825 & \cellcolor[HTML]{63BE7B}0.834 & \multicolumn{1}{c}{1.04\%}  & \cellcolor[HTML]{FDC77D}0.818 & \cellcolor[HTML]{DBE182}0.829 & \multicolumn{1}{c}{1.27\%}  \\ \hline
\end{tabular}}
\end{center}
\end{table*}

The maximum profit for day $d$, called the profit of the \textit{crystal ball} strategy, is given by:
\begin{equation}\label{eq:profit:CB}
\Pi_{\text{CB},d} = 0.9 P_{d,h_2} - \frac{1}{0.9} P_{d,h_1},
\end{equation}
where $h1$ and $h2$ respectively are the hours with the lowest and the highest price of day $d$, and $P_{d,h}$ is the actual price for day $d$ and hour $h$. Clearly, $model$-derived forecasts will yield a profit of
\begin{equation}\label{eq:profit:model}
\Pi_{model,d} = 0.9 P_{d,\widehat{h2}} - \frac{1}{0.9} P_{d,\widehat{h1}} \le \Pi_{\text{CB},d},
\end{equation}
where $\widehat{h1}$ and $\widehat{h2}$ respectively are the hours with the lowest and the highest $model$-predicted prices for day $d$, regardless of the actual price forecasts $\hat{P}_{d,h}$ for this day. In other words, the $model$-derived forecasts will yield a profit equal to that of the crystal ball strategy only if the $model$ correctly predicts the hours with the lowest and highest prices of the day, i.e., if $\widehat{h1}=h1$ and $\widehat{h2}=h2$. 

In Table \ref{table:profits} we report the profits obtained for selected models, expressed as fractions of the profit of the crystal ball strategy for the whole out-of-sample test period, i.e, 
$\Pi_{\text{CB}} = \sum_{d} \Pi_{\text{CB},d}$, 
where the summation is over $d=1.01.2019, 2.01.2019, ..., 31.12.2023$. In our case, $\Pi_{\text{CB}} = 110~897.27$ EUR for EPEX and $\Pi_{\text{CB}} = 55~236.55$ EUR for OMIE. Like in Section \ref{sssec:Temporal}, we only consider models based on the richest pool of individual forecasts (labeled --MAS). The following conclusions can be drawn:
\begin{itemize}
\setlength{\itemsep}{0pt}
\setlength{\parskip}{0pt}
    \item For both the German and Spanish markets, the eSC-type models yield the highest profits within each group (ARX, LEAR).
    \item The best performing model for EPEX is the eSCLEAR-MAS model, while for OMIE the more parsimonious eSCARX-MAS model.
    \item  The differences in profits are more pronounced for the Spanish than for the German market.
    \item On the other hand, the differences in profits are much less pronounced than the differences in predictive performance reported in Tables \ref{table:MAE_results} and \ref{table:RMSE_results}.
\end{itemize}
The latter observation suggests that while the considered models predict price levels much better than the benchmarks, the benchmarks are able to identify the hours with the lowest and highest prices of the day almost as well as the eSC-type models.

\section{Conclusions}
\label{sec:conclusions}

We have introduced a novel approach to predicting the long-term seasonal component (LTSC) for the next day, which is a fundamental input to the seasonal component (SC) approach introduced by \cite{now:wer:16}. Considering parsimonious autoregressive (ARX) and LASSO-estimated autoregressive (LEAR) models, we have provided evidence that improvements in predictive accuracy from using the proposed extrapolated SC-type (i.e., eSC-type) models compared to using a naive prediction of the LTSC can be as high as 15\% for the German EPEX market (in terms of the RMSE; 14\% in terms of the MAE) over a 5-year test period covering the Covid-19 pandemic, the 2021/2022 energy crisis, and the war in Ukraine. 

Interestingly, the improvements with respect to the classical ARX and LEAR models that do not use seasonal decomposition are higher for the Spanish OMIE market and reach up to 10\% (in terms of the MAE; and 8\% in terms of the RMSE); all the differences are statistically significant, as measured by the multivariate variant of the Diebold-Mariano test \citep{zie:wer:18}. Apparently, the models that naively extrapolate the LTSC for the next day do not cope well with the extreme electricity price dynamics in Germany over the period studied and are outperformed by the classical ARX and LEAR benchmarks. 

Furthermore, in line with a recent trend in the electricity price forecasting (EPF) literature \citep{mac:uni:wer:23}, we have considered a realistic trading strategy involving day-ahead bidding and battery storage in order to quantify the benefits in monetary terms. Although for both the German and Spanish markets the eSC-type models yield the highest profits within each group (ARX, LEAR), the differences in profits are much less pronounced than the differences in predictive performance. This may be an indication that while the considered models predict price levels much better than the benchmarks, the benchmarks are able to identify the hours with the lowest and highest prices of the day almost as well as the eSC-type models. 

The latter suggests that computing probabilistic forecasts and considering quantile-based trading strategies \citep{mar:nar:wer:zie:23,nit:wer:23,uni:wer:21} may lead to more significant performance improvements. 
Similarly, studying other energy markets where the seasonal component plays an important role, such as natural gas, may provide evidence that the proposed approach extends beyond electricity markets. Investigating the impact of seasonal decomposition on other machine learning models, including deep neural networks, may open new avenues of research. Finally, although the use of simple averaging was effective in this study, more sophisticated ensembling techniques can be considered. All this, however, is left for future research.

Overall, our results highlight the importance of seasonal decomposition and accurate day-ahead forecasting of the trend-seasonal pattern of electricity prices. The proposed approach is robust and ensures good performance even under extremely volatile market conditions. Due to the use of forecast averaging, it does not require ex-ante selection of the LTSC parameters (width of the moving average, wavelet decomposition level). 
Its simplicity and low computational requirements make it a perfect tool for daily market operations, both for point forecasting tasks and as a reliable source of inputs for probabilistic forecasting models and risk management applications \citep{gne:ler:sch:23,lip:uni:wer:24}.

\section*{Acknowledgments}
This work was partially supported by the National Science Center (NCN, Poland) through grants No.\ 2018/30/A/HS4/00444 (to K.C.), No.\ 2023/49/N/HS4/02741 (to B.U.) and No.\ \linebreak 2021/43/I/HS4/02578 (to R.W.).


\bibliography{lasso}

\end{document}